# Visualising highly localised luminescence in GaN/AlN heterostructures in nanowires

L. F. Zagonel[1,2], L. Rigutti[3], M. Tchernycheva[3], G. Jacopin[3], R. Songmuang[4], M. Kociak[2]

1. Laboratório Nacional de Nanotecnologia, CNPEM, 13083-970, Campinas, Brazil

2. Laboratoire de Physique des Solides, UMR CNRS 8502, Université Paris Sud 11, 91405 Orsay Cedex, France

3. Institut d'Electronique Fondamentale UMR CNRS 8622, Université Paris Sud 11, 91405 Orsay Cedex, France

4. CEA-CNRS group "Nanophysique et Semiconducteurs", Institute Neel, 25 Rue des Martyrs, 38054, Grenoble cedex 9, France

*Abstract*

The optical properties of a stack of GaN/AlN quantum discs (QDiscs) in a GaN nanowire have been studied by spatially resolved cathodoluminescence (CL) at the nanoscale (nanoCL) using a Scanning Transmission Electron Microscope (STEM) operating in spectrum imaging mode. For the electron beam excitation in the QDisc region, the luminescence signal is highly localized with spatial extension as low as 5 nm due to the high band gap difference between GaN and AlN. This allows for the discrimination between the emission of neighbouring QDiscs and for evidencing the presence of lateral inclusions, about 3 nm thick and 20 nm long rods (quantum rods, QRods), grown unintentionally on the nanowire sidewalls. These structures, also observed by STEM dark-field imaging, are proven to be optically active in nanoCL, emitting at similar, but usually shorter, wavelengths with respect to most QDiscs.

*Published on IOP Nanotechnology*

1. Introduction

Complex nanostructured materials, such as heterostructured semiconductor nanowires (NWs), are being extensively studied due to their promising properties for the development of novel nanoscale devices.[1,2,3] III-Nitride NWs are particularly attractive for photonic applications because of the nitride direct band gap ranging from the deep UV (6.2 eV) for AlN to near infrared (0.7eV) for InN. They can be grown by different techniques and assembled into complex devices [4,5,6,7,8,9]. The study of their optical properties requires characterization tools with nanometric spatial resolution [10]. All-optical techniques, like photoluminescence or near field optical microscopy, fail to go far below the diffraction limit and to reach a spatial resolution near 1 nm. Cathodoluminescence (CL) relies on electron probes, which can be nowadays smaller than 100 pm, and hence provides high-resolution luminescence maps [11,12,13,14]. Nevertheless, the resulting luminescence is usually localized on a much larger scale.[15,16] In bulk materials, indeed, the electron probe spreads over distances of the order of tens to hundreds of nm. In the case of NWs, even if the interaction volume is strongly reduced, the generated electron-hole pairs may diffuse several hundreds of nanometres before eventually recombining. Hence, the ultimate spatial resolution in cathodoluminescence experiments is in fact sample dependent. This feature can be used advantageously in the study of confinement effects and effective carrier diffusion lengths in NW heterostructures [17]. Several CL studies of the optical properties of III-N heterostructured NWs have recently appeared [18, 19, 8, 20]. In particular, our group has studied the Quantum Confined Stark Effect (QCSE) for each individual GaN/AlN QDisc of width known with atomic precision in NWs containing multi-QDiscs stacks.[11] Similarly, we have recently observed individual Nitrogen-Vacancy centres in diamond with high spatial resolution and InGaN inclusions in GaN NWs.[21, 22]

Here we report a study of the cathodoluminescence of GaN NWs containing a stack of GaN QDiscs of different thicknesses surrounded by AlN barriers. Using CL at the nanoscale (nanoCL) spectral imaging in a Scanning Transmission Electron Microscope (STEM), we aim on finding the nature and emission energy of optically active features in the heterostructure. We observe that the cathodoluminescence signal of GaN quantum structures within AlN barriers can be as local as 5 nm (signal spatial extension is defined as a region outside which the CL intensity decreases by a factor of e from emission maximum). The simultaneous acquisition of the high-angle annular dark field (HAADF) STEM image with the CL signal allows attributing the spatially resolved CL emissions to specific QDiscs in the stack. It also allows evidencing the luminescence of lateral inclusions, called here Quantum Rods (QRods), unintentionally formed on the lateral NW sidewalls.

2. Experimental details

The studied NWs have been grown on Si(111) substrates at 790C by using radio frequency plasma assisted molecular beam epitaxy (PA-MBE) without using any catalysts [23]. The NWs consist nominally of half micron long GaN base followed by a stack of 20 AlN/GaN QDiscs. The QDiscs are formed by switching from Ga to Al flux without growth interruption. The heterostructure region is then followed by another half micron GaN part [7]. The GaN extremities of the NW as well as the GaN QDiscs are doped with Si at a nominal density $N_d=10^{19}cm^{-3}$. Due to the AlN lateral growth, the QDisc region and the GaN base are surrounded by an external AlN shell, while the top GaN part remains uncovered. [24, 25, 26] An additional partial GaN shell can be formed during the growth of the top GaN segment and for some NWs lateral inclusions of GaN in the AlN shell also appear [26].

NanoCL imaging was performed in an optimized system with maximized collection angle and high spectral resolution.[27] The system consists of an optical spectrometer and a CCD camera for parallel spectra acquisition. The nanoCL device is installed in a VG 501 HB STEM, equipped with a high brightness cold field emission gun and a specially designed scanning module that is triggered by the CCD camera. The STEM was operated at 60 kV with a probe current of about 300 pA. The sample stage was cooled down to about 150 K with liquid nitrogen and the TEM grid was firmly attached by use of silver paste. The acquisition was done in the spectrum imaging mode, i.e. for each pixel of the image, a full CL spectrum is recorded. We obtain in this way a 3D ($I_{CL}$ (x,y, λ)) dataset (usually called datacube), with x,y the spatial coordinates, $I_{CL}$ the cathodoluminescence intensity, and λ the wavelength. Figure 1 illustrates the acquisition scheme. The electron beam scans the sample and for each position both the Dark Field Detector signal and the CL spectra are recorded. At the end of the scan, both a Dark field image – giving structural information – and a CL spectral image are generated. Both spectral and structural information can in this way be correlated pixel per pixel. In Figure 1, the spectrum Image principle is exemplified in the case of a nanowire. For this nanowire, both an opaque volume containing the CL datacube acquired and the DF image (shown on top) are displayed. In this representation, low CL intensity (i.e.: background) is set to be transparent while higher CL intensity is set to a colour scale from violet to white. In the Spectrum Image, it is possible to see the strong emission from the GaN bulk covered by AlN on the right-hand side (intercepting the vertical plane on the right) and an isolated feature near the volume centre. This feature is isolated spatially and spectrally in this view while by rotating this volume the isolated nature of some other features can be unveiled. Typical dwell times per pixel were 20 ms, the spatial sampling was about 1 nm and the spectral resolution was 1 nm. High resolution (HR) TEM images were acquired with a Nion UltraSTEM 200 featuring a $5^{th}$ order aberration corrector.

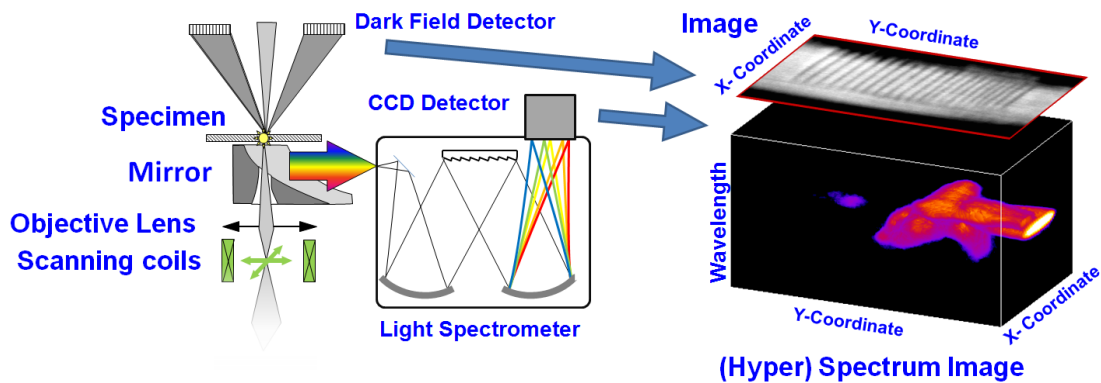

Figure 1: Acquisition scheme for spectrum imaging. On the left, in the microscope (STEM) two kinds of detectors are used: a Dark Field Detector and a CCD camera attached at the exit of an optical spectrometer. Upon the electron beam scanning on the sample, the former will produce an image while the later, having collected a spectrum at each pixel of such image, will provide a Spectral Image. On the right, one DF image is shown together with an illustrative view of a Spectrum Image. In the latter, the light emission is shown as an opaque volume. Within this volume, it is possible to distinguish an isolated emission of the middle.

3. Results

Figure 2 reports the HAADF STEM images of the QDiscs region of four NWs from the same sample grown in the conditions described in section 2. The QDiscs are visible in the middle and the GaN top and base are shown on the left and right side of the images, respectively. Figure 2(a) shows a NW with a well-defined QDisc system. The heterostructure region consists of GaN QDiscs with thickness progressively increasing from 1 to 4 nm in the growth direction, while the AlN barrier thickness is about 4 nm. A lateral AlN shell surrounds the QDisc region and the GaN base. The thickness of the AlN lateral shell gradually decreased from 5-10 nm at the bottom of the QDisc region and around the GaN base to at the top of the QDisc region. No shell is present on the GaN top part. Such nanowires are the most frequent, corresponding to the expected growth morphology, as previously reported [11]. The nanowire depicted in Figure 2 (b) is similar to the previous one, except for the presence of many small GaN inclusions on the sides of the QDisc stack. Similarly, the HR-STEM image of a NW with inclusions in both sides is shown in Figure 2(c). A sketch of these inclusions is shown in Figure 2(d). These inclusions have already been observed in PA-MBE grown GaN/AlN NWs and the mechanism leading to their formation is discussed in ref. [28]. The inclusions appear unintentionally during the growth on either one or several NW facets. They have a rod-like shape, i.e. they have two short dimensions in the image plane (of the order of 2-3 nm), but their extension along the beam direction is equal to the NW facet width, about 20 nm. This third dimension is larger than the exciton Bohr radius, therefore these rod-like nanostructures provide quantum confinement only in 2 dimensions. Thus we will refer to them as to quantum rods (QRods), as their shape is similar to that of QRods in the colloidal II-VI materials systems [29,30]. It should be noted that, contrary to the case of the QDiscs, in order to visualize the lateral QRods, it is necessary to orient the NW so that the particular facet presenting the rod overgrowth is parallel to the electron beam. This usually requires the possibility of tilting the sample by typically ±30°. Since the tilt range of the sample holder in the used STEMs is limited to 0 for Cathodoluminescence, and ±25° for HR imaging, this is not always possible. Figure 2(e) shows an example of a NW with a facet not aligned with the electron beam, for which the QRods cannot be visualized. Indeed, the image shows only a large bright region on one side of the QDisc system. By tilting the sample, it is possible to orient the NW facet with respect to the beam and then observe the QRods, as shown in Figure 1(f). NanoCL mapping was performed on the NW shown in Figures 1(e) and (f).

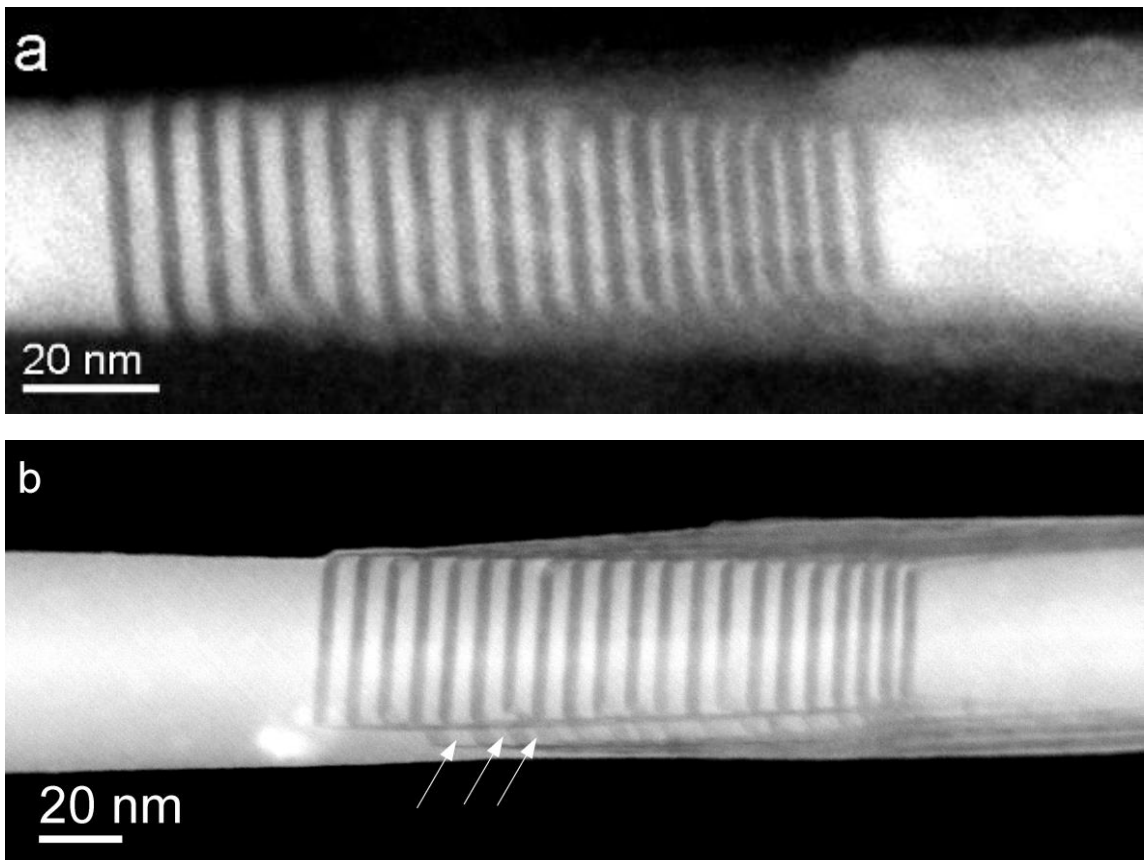

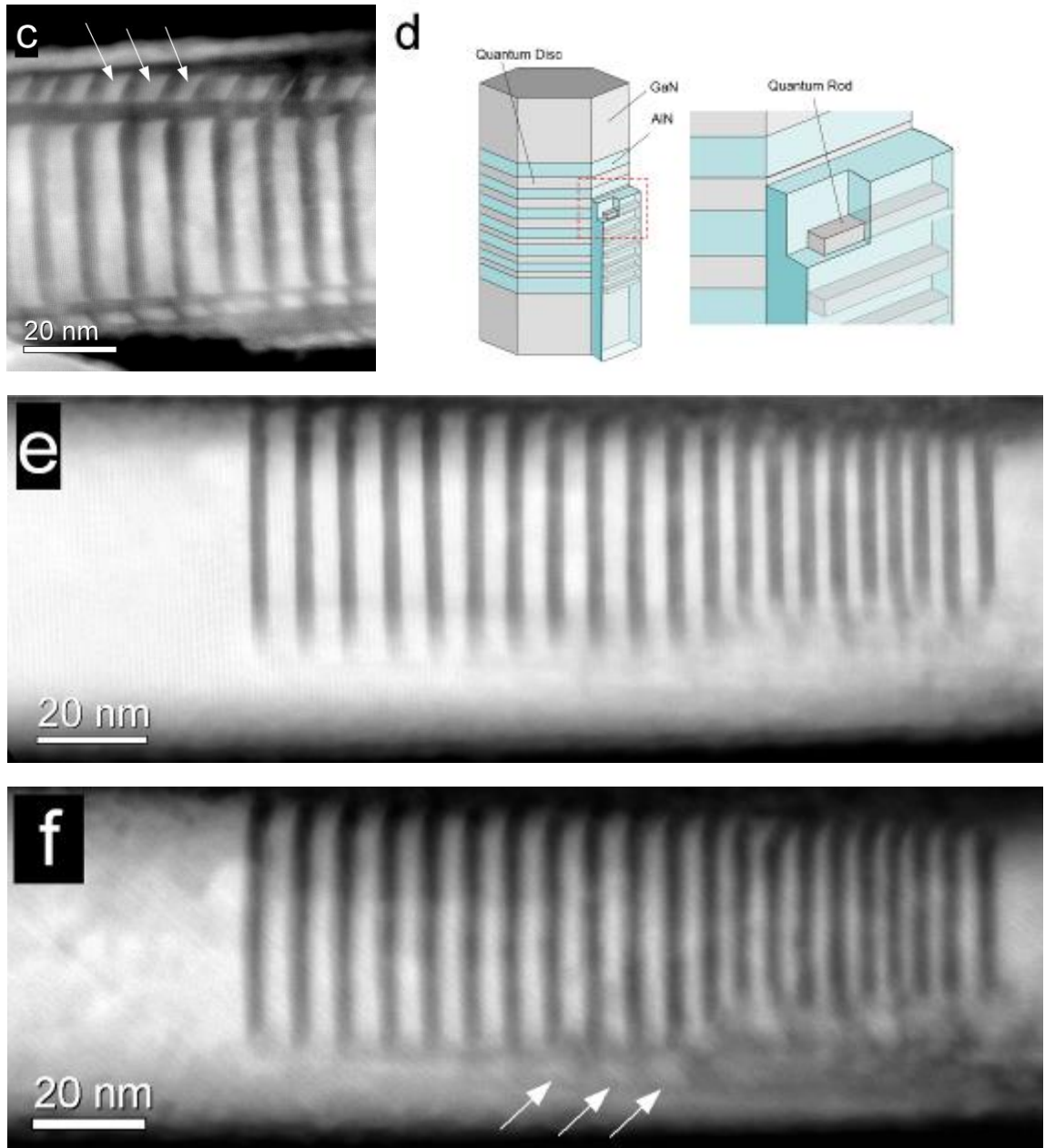

Figure 2: STEM-HAADF images of three different NWs from the same sample. (a) A NW with only a well-defined QDisc system. (b) A NW with QRods GaN/AlN structures on the side of the QDisc stack. (c) A close-up in a NW with QRods on both sides. In both case, some QRods are indicated by arrows. (d) A sketch of the NW with some QDiscs and the side QRods, similar to the one shown in (b). (e) A NW oriented with a facet perpendicular to the electron beam (in this orientation the QRods are not visible). (f) The same NW as in (e) in other orientation (tilted around the growth axis) with QRods clearly visible on the bottom of the image (indicated by arrows). The same NW as in (e) and (f) is shown in Figure 4(a). The growth direction is from right to left in all figures.

In order to analyse the recorded CL spectrum images it is necessary to work with both spatial and spectral dimensions simultaneously. To illustrate this, the Figure 3 shows different ways to inspect a spectrum image looking for the presence of an optical signal (labelled by a well defined peak wavelength in the following cases) coming from a small region on the sample. First, slices of the spectrum image having 2 spatial coordinates at a given wavelength (usually refered to as a wavelength filtered CL image, or simply a CL image) show the intensity of light emission at such wavelength (top-left panel). A small feature in the CL image is observed when filtering at 318 nm. Other signals can be seen to appear at similar wavelengths, or even be higher in intensity at the same spatial location but shifted in wavelength, thus hiding the signal of interest at a particular location. When aiming at tracking the physical location of the emission, it is thus necessary to consider the spatio-spectral variation of this signal all together. As the full 3D datacube is difficult to visualize, we analyse slices from the spectrum image along two planes mixing both the spatial and spectral information: one slice is ploting the x-coordinate vs. wavelength (for a given y-position) and the other is plotting the y-coordinate vs. wavelength (for a given x-position). These are also shown in Figure 3. In each one, the correlations between a particular spectral component and a particular position (along x or y) can be analysed. We clearly see that the 318 nm peak is spectrally *and* spatially well separated from the rest of the emission in *both* the x and y directions. Such an observation holds for any other slice orientation (not shown) for which one direction is the spectral one. The spatial position at which the 318 nm peak is maximum on the two slices is obviously the same as that revealed in the (318 nm) CL image. It is also possible to see that at the same position there are emissions at other (longer) wavelengths coresponding to other optical active strucutres (like QDiscs or others QRods). This is normal since all features along the electron axis can be excited: the electrons goes all their way through the sample. Nevertheless, the detailed spectral information allows us to distinguish different optical active features by their *combined* wavelength and position maxima in the full spatio-spectral space. Finally, by summing all spectra from the region of the feature of interest (around the spatial positions where the 318 nm signal shows up), one obtains a spectrum, with all contribution from that region, that shows a peak at this specific wavelength (indicated by a dashed circle in the bottom-right panel of Figure 3). As the CL maps alone, the spectrum alone cannot be an absolute proof of the presence of an emitting QRod. This is especially true in this example where the QRods emission is particularly weak with respect to the other emisions, but can be easily retrieved through spectral imaging analysis. This analysis was performed for other optically active features and the results for another QRod are shown in Figure S1 in the Supplemental Material. In the sequence of the analysis, we will show only CL maps and spectra (like the top-left and bottom-right panels in Figure 3, respectively), but this procedure for isolating localized signal (in space and spectra) has been sucessfully performed for each of the identified features.

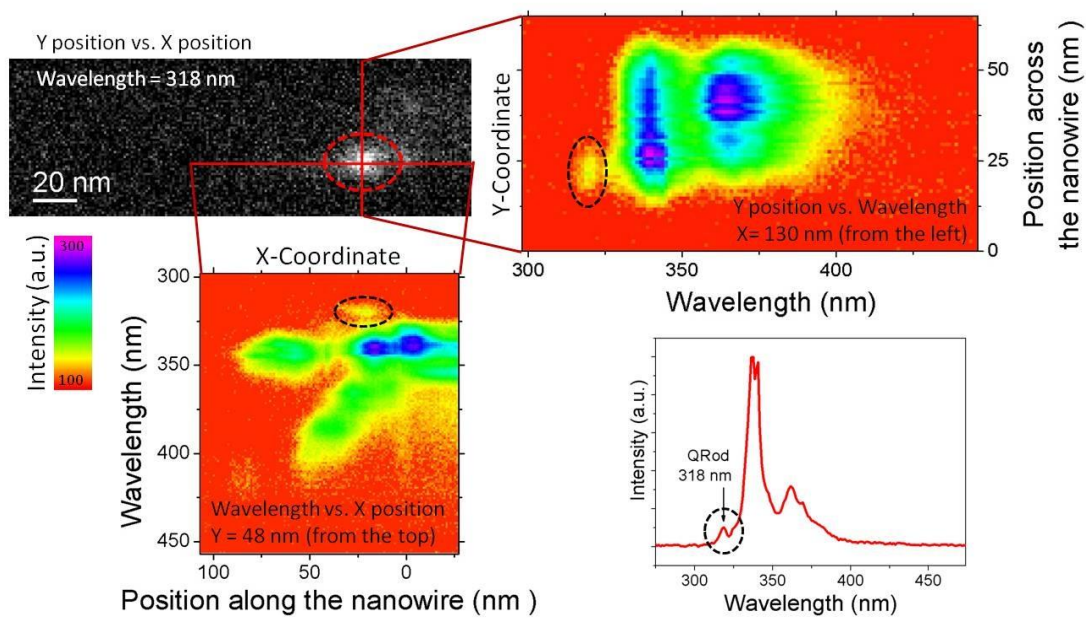

Figure 3: The identification of highly localized optical signal depends upon the analysis of slices and integrated sub-volumes extracted from the spectrum image. Top-left (2 spatial coordinates at fixed wavelength): typical wavelength filtered CL image within a narrow wavelength window shows optical activity at 318±4 nm. On top-right and bottom-left (1 spatial and 1 spectral coordinates from a given position): slices from the spectrum image along the spectral direction and two different spatial direction (paralel and perpendicular to the x axis). In both panels, the feature observed in the filtered CL image (dashed red circle) is observed to be separated spectrally and spatially from other signals (dashed black circles). Bottom-right: by summing all spectra from the region of interest (red dasehd circle in the CL image) the spectrum from the feature is obtained (peaked in 318 nm and marked by a black dash circle).

An image of the NW with the QDiscs stack acquired just before the NanoCL mapping is shown in Figure 4(a). CL maps $(x,y,I_{CL})$ for fixed detection wavelengths extracted from a typical spectrum image of GaN QDisc region are shown in Figure 4 (b) to (g); see the full spectrum image in Movie M1 of Supplementary data. The emission wavelengths are shorter, equal and longer than the emission from the bulk GaN, which is 357 nm. In Figure 4(b) and (c) we observe emissions, marked with circles, localized outside the region of the QDiscs on the side of the NW. These emissions are peaked at 318 nm (in Figure 4(b)) and at 336 and 341 nm (left and right, in Figure 4(c)). As shown in Figure 3 for the feature at 318 nm, the other 2 are also isolated spectrally and spatially (the feature on the left on Figure 4(c) is the same one observed in the middle of the Spectrum Image volume plot in Figure 1 and the feature on the right on Figure 4(c) is shown in detail in Figure S1). Due to their location outside the QDisc stack, we attribute these peaks to the QRods grown on the lateral facets, as discussed above, in agreement with the morphology of this NW (shown in Figure 2(f)). Indeed, although usually the QRods are not seen in nanowires without tilting the sample (which is not compatible with our NanoCL set up), they are optically active and therefore can be evidenced in CL maps. More, the same NW can be *a posteriori* tilted in a tilting sample holder to evidence the presence of inclusions at the position of lateral emission (Fig.2 (f)). The spectra corresponding to these regions are displayed in Fig. 5(c). The short wavelength of the luminescence of some QRods with respect to that of most of the QDiscs is possibly related to their size, smaller than that of the QDiscs (Fig. 2(b,c and f)), and to the quantum confinement in two dimensions.

It is important to note, however, that, particularly for the emission indicated by the circle in the right on Figure 4(c), it is not possible to be certain of the exact morphology of the optically active features. This is because the HAADF image acquired simultaneously with the CL spectrum image does not reveal the morphology of the QRods (even if they are present as evidenced in Figure 2(f)) and, even by a carefully tilting the sample, the morphology of some features remain unclear. Indeed, a full morphological analysis would require high resolution tomography which is out of the scope of this study. Therefore, one must be open to the possibility that optically active features grown on the side facets might differ from the concept indicated on Figure 2(d). Nevertheless, in all NW observed, only features similar to those shown in Figures 2 (b) and (c) were observed.

In Figures 4 (c) to (g), the emission from different single QDiscs can be distinguished. Along the NW axial direction, the spatial extension of the CL signal can be observed in these images to be of the order as the distance between adjacent QDiscs (measured as 3 to 4 nm in average using HR-STEM). Indeed, the intensity at specific wavelengths is concentrated on small regions, particularly over a single QDisc. Note that the emission is not observed from all QDiscs, possibly due to the presence of non-radiative point defects in the QDiscs or defects in the AlN barriers and shell. The QDiscs located closer to the NW base (right hand side in Figure 4) are brighter than those located closer to the top (left hand side in Figure 4). As for the GaN base, the AlN overlayer around QDiscs acts as a barrier preventing carriers from diffusing to the surface where non-radiative recombination is likely to occur.[31] We also note that the QDiscs emission wavelength shifts to the red for QDiscs located closer to the NW top. This shift is induced by the increase of the QDisc size in the growth direction amplified by the QCSE. In addition, the AlN shell gets thinner towards the NW top. As a result, the strain is weaker for the upper QDiscs, which reduces the strain-induced blueshift [11,32].

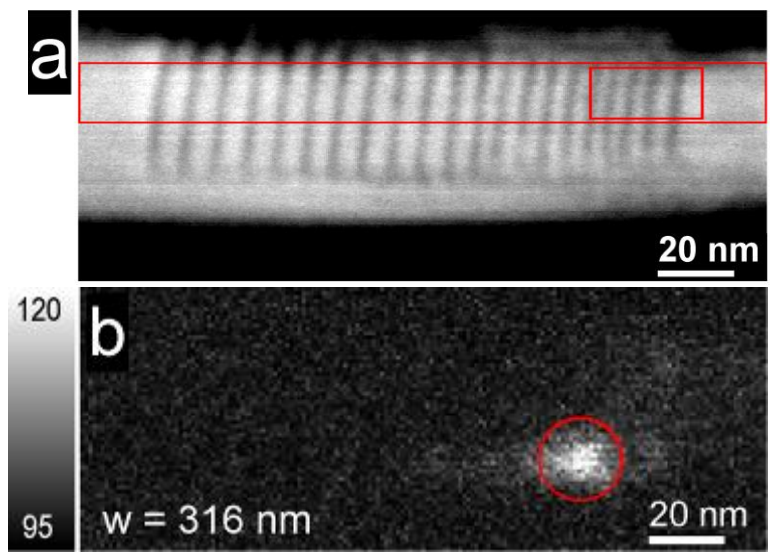

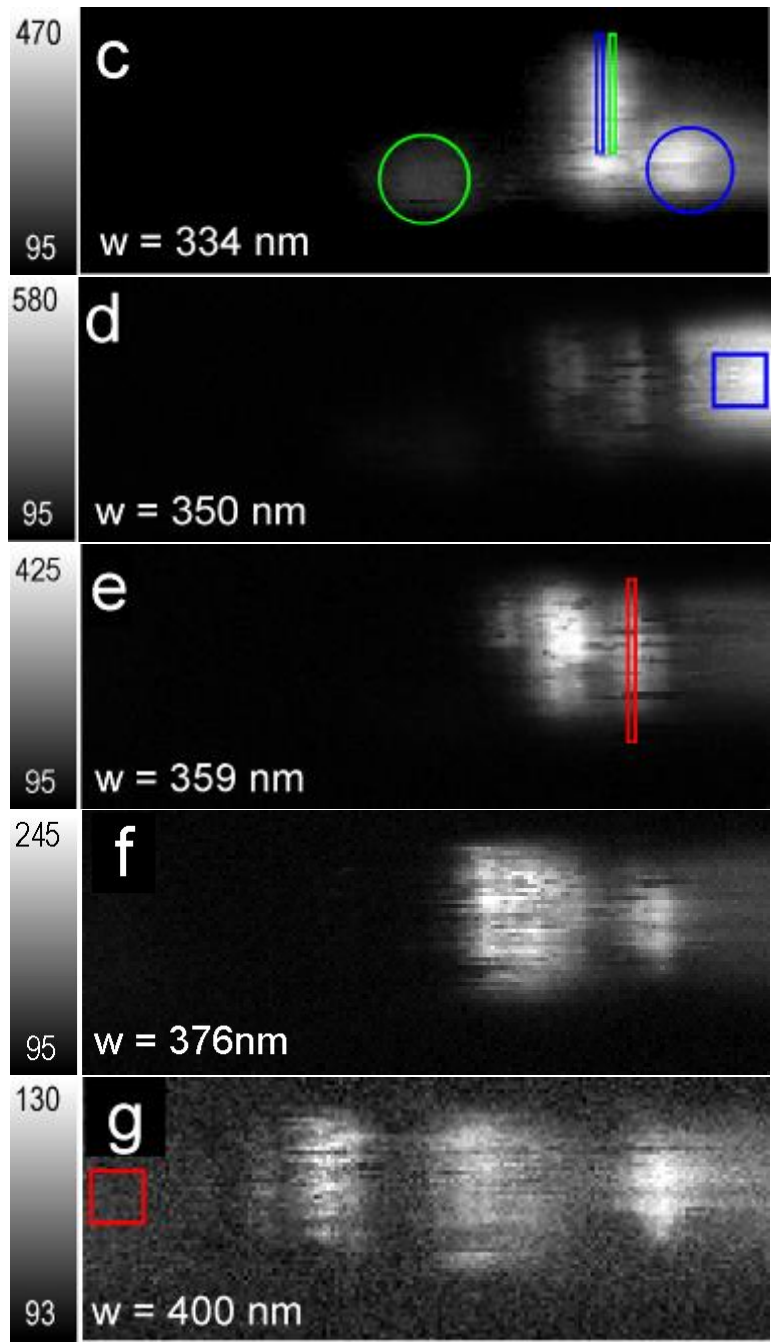

Figure 4: (a) HAADF STEM image of the QDisc stack in the middle of the NW acquired just before the NanoCL mapping. Growth direction is from right to left. The rectangles show the regions from which the profile of Figure 6 (inner) and projection in Figure S2 (outer) were extracted. (b) to (g) show energy filtered CL images obtained from the spectrum-image at wavelength of 316, 334, 350, 359, 376 and 400 nm, respectively (spectral integration range is ±4 nm). In these images, the emission of single QDiscs can be distinguished. The circles in (a), (b) and (c) show the emission attributed to QRods on the side facet of the NW. Rectangles in (c), (d), (e) and (g) show the emission from QDiscs and the GaN bulk on the base and top parts of the NW. These regions are indicative positions of the extracted spectra for Figure 5. The field of view in all figures is the same. Due to sample drift during acquisition, the aspect ratio of these images is compressed by 20% in the vertical direction with respect to the real object (Figure 2(e) shows NW morphology with balanced aspect ratio).

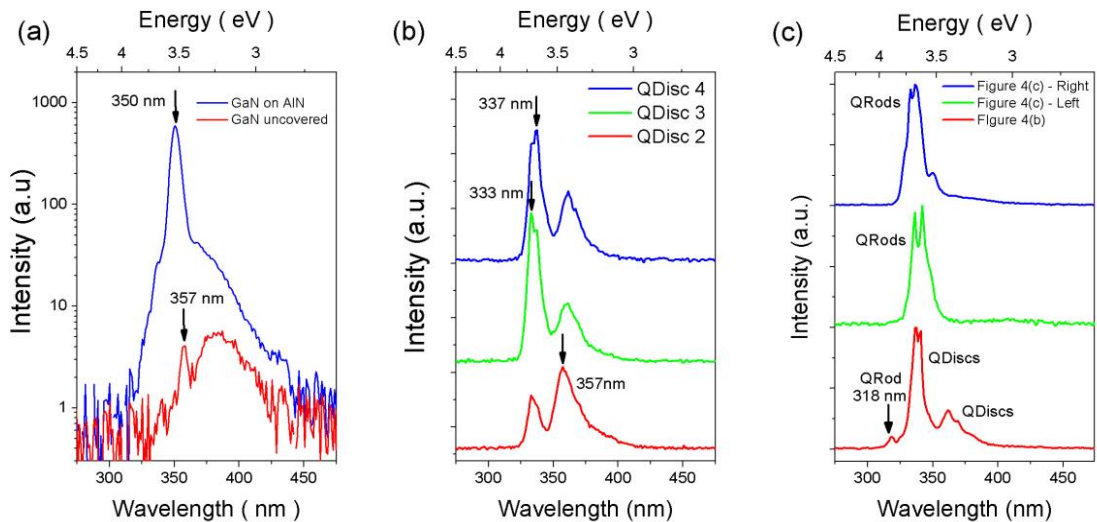

Figure 5: Spectra extracted from the spectrum image at different positions. (a) Parts of the GaN NW either covered with the AlN shell or not (the spectra are extracted from the regions marked with squares in Fig. 4(d) and (g)). Note the logarithmic scale. (b) Selected QDiscs (the spectra are extracted from the regions marked with rectangles in Fig. 4(c and e)). (c) Region of lateral inclusions attributed to QRods (the spectra are extracted from the region marked with circles in Fig. 4(b and c)). The vertical arrows and the labels indicate the emission peak wavelength.

The NanoCL map recorded at a wavelength of 350 nm (Fig. 4(d)) shows the emission from the GaN NW base part (on the right side) while the corresponding spectrum is displayed in Fig. 5(a). The GaN base emits at λ=350 nm, at an energy slightly higher than the bandgap of bulk GaN at 150K, due to the strain exerted on it by the AlN shell [33]. The GaN side not covered by AlN emits at 357 nm (but the signal is 150 times weaker) together with a broad emission at 384 nm, possibly related to defects [34,35]. This is most likely due to the presence of surface non-radiative recombination channels, which may be effectively passivated by the AlN layer. This passivation effect has been demonstrated by the quenching of the 3.45 eV PL emission line in GaN nanowires, which does not appear in nanowires covered by AlN [33]. We note that there was no evidence in our experiments of the presence of the Yellow Band (YL), related to defects, usually appearing at about 560 nm.[36,20] This indicates the good crystal quality of the NWs used in this study. Moreover, differences in radiative lifetimes for near band edge luminescence and YL may affect our perception of the YL, particularly since this signal can have much lower intensity. [37,38,39]

The spectra corresponding to the position of each QDisc can be extracted from the spectrum image and some are reported in Fig. 5(b) for the 2nd, 3rd and 4th QDiscs. Using Gaussian fit, the peak wavelength of each spectrum is extracted in correspondence to the emission of individual QDiscs, i.e. λ=357, λ=333 and λ=337 nm, respectively. Here again, the proper identification of the emission energy of each QDisc is performed in spectral-spatial plots, as shown in Figure 3 and Figure S2 in the Supplementary data. The thickness of each QDiscs, measured with monolayer (ML) resolution in HR-HAADF-STEM images, is respectively, 8 ML, 6 ML and 6 ML. The 5th QDisc, emitting at 359.5 nm, has 9 ML. The relation between the emission wavelength of each QDisc and its thickness is in perfect agreement with those measured previously [11]. The spectral broadening of each individual emission is about 10 nm.

The spectra corresponding to the emissions observed on the side of the NW, attributed to QRods, are shown in Fig. 5(c). It is observed that the emissions from the regions shown in Fig. 4(c) have more than one peak what could indicate the emission of adjacent QRods in each of those areas. On the contrary, the emission around 318 nm (Figure 4(b) and Figure 3) corresponds to a single peak, even if it is very small with respect to the surrounding QDiscs. The luminescence signals that peak in the regions noted by circles in Fig. 4 can be readily associated to QRods (i.e.: the inclusions on the side of the NW). Indeed, while all observed QDiscs have their emission spread in the section perpendicular to the NW growth direction, the emission associated to QRods, on the other hand, are concentrated in a small region on the side on the NW. It should be noted that QRods can also be present on the NW facets parallel to the image plane. However, their emission, if present, cannot be unambiguously distinguished from that of the QDiscs. This would require tilting the sample while performing CL measurement, which is actually not possible.

Concerning the luminescence of the lateral QRods, we can also suspect that not all of them are optically active. Indeed, only about 5-7 emissions from side QRods are observed in nanoCL maps of Figure 4 (see also Movie M1 in Supplementary data), although in the STEM images many QRods are observed all along the NW side (Figure 2). The lateral QRods more likely suffer from non-radiative recombinations than QDiscs since they are separated from the surface by a thinner AlN layer than the QDiscs. In addition, because of the confinement in two dimensions, the recombination dynamics is more sensitive to the crystalline quality of the AlN barriers. Finally, because of the lattice mismatch in the core-shell AlN/GaN heterostructure, the AlN shell may present dislocations perpendicular to the NW axis [40], that we could not be evidenced in the present case.

As we have mentioned previously, the cathodoluminescence is highly localized, with a typical spatial signal extension of the same order as the distance between adjacent QDiscs. To determine quantitatively the spatial extension of the luminescence, one can analyse intensity profiles from CL maps at a fixed detection wavelength which originating from a single quantum object. For clarity's sake, we only show here the profile analysis of the emission of the 3rd and 4th QDiscs along the NW growth axis. These intensity profiles are reported in Figure 6 (See details in the Supplementary data, Figure S2). The profiles give the emission intensity of an individual QDisc for different locations of the electron probe: from the QDisc position to some tens of nm away. As shown in Figure 3, the emission of a given QDisc is distinguished by its wavelength. The profile, in Figure 6, is flat inside the QDisc and decays steadily outside. When the probe hits the QDisc directly, the emission is at its maximum. As the electron probe distance from the QDisc centre increases, the CL signal intensity decreases. The CL profiles of the 3rd and 4th QDiscs can be fitted with an exponential curve with a characteristic length of 5±1 nm. This value is about the same as obtained in a global fit to the spectrum image shown in the Supplementary data (Figure S2). These fits yield an estimation of the signal localization across the sequence of GaN and AlN layers. Moreover, laterally, the signal peaks at the center of the QDisc, and drops slightly quicker than the HAADF signal with a vanishingly small contribution outside of the NW, indicating a bulk nature of the excitation process, in contrast to plasmons excitations in nanoparticles of rather similar sizes [41].

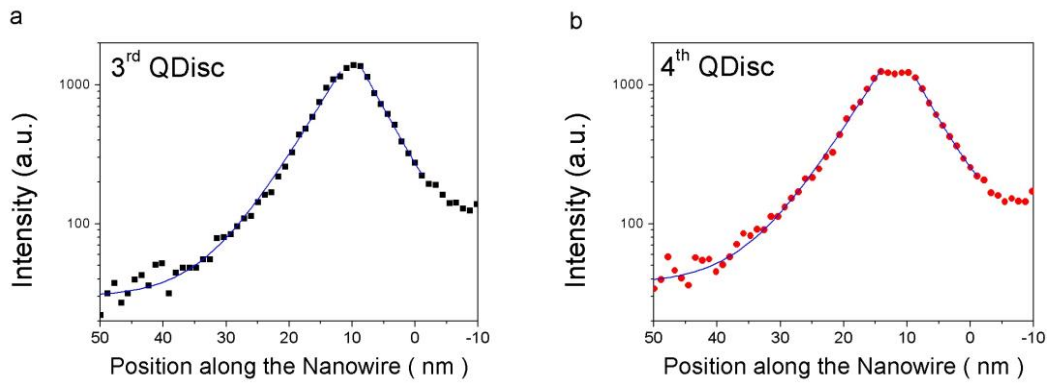

Figure 6: Dependence of the CL signal intensity at λ=333 nm (a) (coming mostly from the 3[rd] QDisc) and at λ=337 nm (b) (coming mostly from 4[th] QDisc) on the electron probe position (see horizontal lines in Figure S2(b)). The intensity is higher on the right-hand side of the graph due to the strong background coming from the bulk GaN emission at λ=350 nm.

The extension of the NanoCL signal is determined by different physical phenomena discussed below. It is well known that the electron can create an electromagnetic field that can be absorbed by the material in the form of surface or bulk excitations thanks to Coulombic coupling. In semi-conductors, if these excitations directly produce an electron-hole pair, or can be converted to electron-hole pairs, they might be the source of luminescence. For a bulk excitation, the localization is supposed to be less than one nm. Surface excitations can extend over larger distances (few nanometers) [41]. However, it is obvious that if surface excitations plays a role, their contribution is much smaller than the bulk one for nanowires (this is for example already the case for an ultrathin and empty object such as a carbon nanotube [42], and even more so for semiconducting nanowires [43]). It must be noted also that, even if the Coulomb interaction reaches up to roughly the electron speed divided by the excitation frequency, the electric field is very strong for very short distances and hence high contrasts are to be expected between the actual electron path and a region few nm away.[12] Another source of delocalization, secondary electron enhancement of the CL signal [44], is also to be ruled out, as emission of QDiscs from one wire isn't triggered when the electron beam passes on an adjacent wire when studying NW bundles. A charge carrier diffusion mechanism should therefore be the origin of the signal spatial delocalization. However, the effective diffusion length is very short due to the multiple barriers in the GaN\AlN heterostructures.[14]

Indeed, for AlN or GaN, the diffusion length, usually measured for homogeneous materials, is much longer than the CL signal spatial distribution observed here. For comparison, the carrier diffusion length observed in GaN NWs or high quality bulk GaN is in the range of ~1-3 μm [45,46]. Other reports give shorter diffusion lengths in the range of 10 to 100 nm, due to the presence of defects [47,48,49]. The signal extension observed in the present study is much smaller than all reported values. Such short diffusion length cannot be explained only by defects or high density of non-radiative centres, since their concentration is low in NW material. Therefore it should be due to the band structure of the AlN/GaN stack. Indeed, the QDiscs can efficiently capture the excited electron-hole pairs and the high confining potential of AlN barriers prevents their further diffusion. Also, in this heterostructure high internal electrical fields may lower the electron-hole overlap everywhere but within the discs where high radiation efficiency is to be expected [11].

4. Conclusions

In conclusion, the optical properties of the GaN/AlN QDiscs in nanowires have been probed by low temperature nano-cathodoluminescence in a STEM. The coupled structural and optical studies allowed to evidence the presence of optically active QRods formed on the lateral nanowire facets. The luminescence signal in axial GaN/AlN heterostructures is localized on typical distances as short as 5±1 nm, more than two orders of magnitude shorter than the carrier diffusion length in bulk GaN. This dramatic reduction is due to the high confining energy of AlN barriers and to the high recombination probability inside each GaN confined nanostructure (i.e. QDiscs and QRods).

**Acknowledgements:** We acknowledge useful discussions and support from O. Stéphan and C. Colliex. The authors acknowledge financial support from the European Union under the Framework 6 program under a contract for an Integrated Infrastructure Initiative. Reference 026019 ESTEEM. The research described here has been partly supported by Triangle de la physique contract 2009-066T-eLight and by the French ANR agency under the program ANR-08-NANO-031 BoNaFO. This work has received support from the National Agency for Research under the program of future investment TEMPOS-CHROMATEM with the reference ANR-10-EQPX-50.

# Supplementary data
## Video

The data shown in Figure 1, 3 and 4 of the main text has been extracted from a whole spectrum image, which has 256 images and is shown in the video V1.

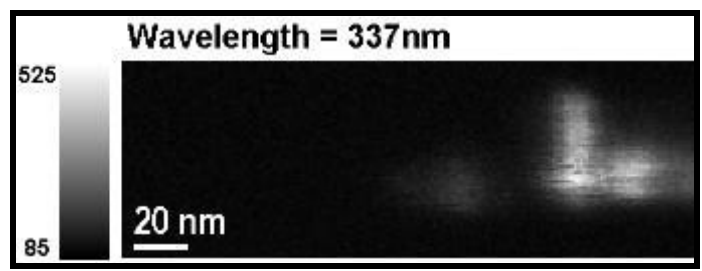

Video: Spectrum Image showing the emission from the GaN QDiscs and QRods.

The analysis shown in Figure 3 of the main text can be done for other optically active features. The QRod of the right hand side of Figure 4(c) is shown in details in Figure S1. Similarly as in Figure 3, the panels show different information extracted from the same spectrum image shown in the Video. The panel on the top-left shows an energy filtered CL image that displays the high intensity emission at a given spot on the side of the NW. By analysing spatial-spectral plots for this position it is observed that indeed, a feature spectrally and spatially isolated is emitting from the side of the NW at about 340 nm. This feature can be attributed to a QRod given the spatial distribution of this emission.

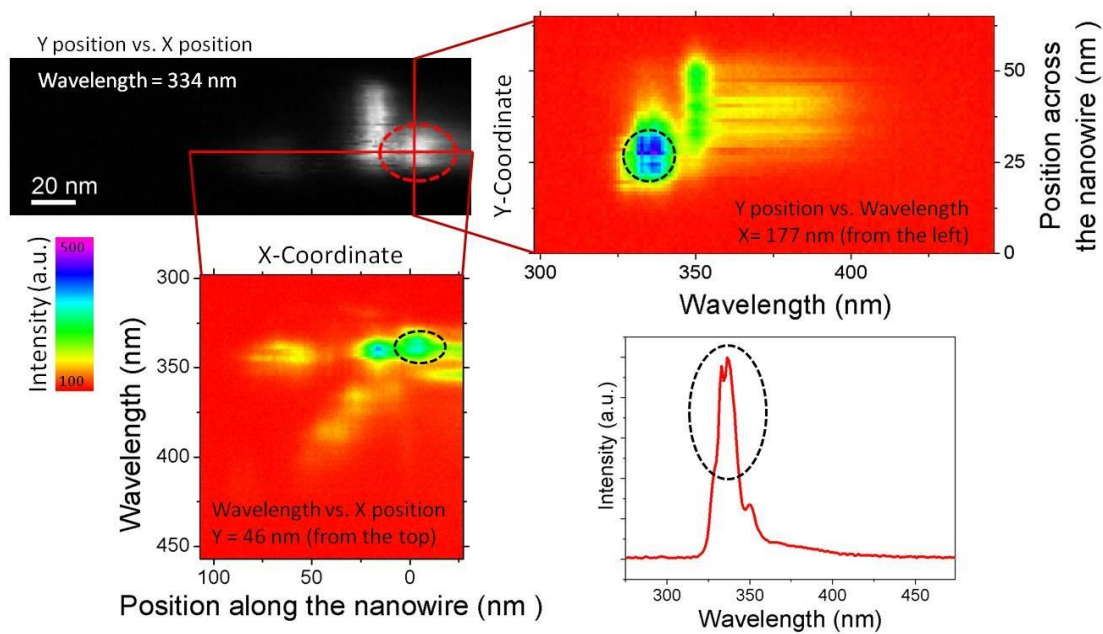

Figure S1: The identification of highly localized optical signal depends upon the analysis of slices and integrated sub-volumes extracted from the spectrum image. Top-left (2 spatial coordinates at fixed wavelength): typical wavelength filtered CL image within a narrow wavelength window shows optical activity at 334±4 nm. On top-right and bottom-left (1 spatial and 1 spectral coordinates from a given position): slices from the spectrum image along the spectral direction and two different spatial direction (paralel and perpendicular to the x axis). In both panels, the feature observed in the filtered CL image (dashed red circle) is observed to be separated spectrally and spatially from other signals (dashed black circles). Bottom-right: by summing all spectra from the region of interest (red dasehd circle in the CL image) the spectrum from the feature is obtained (peaked in 336 nm and marked by a black dash circle).

A more global way to investigate the light emission from the QDisc stack is to observe the spread of the signal of each QDisc on the surroundings in a combined spatial-spectral plot, as shown in Fig. S2. Fig. S2(a) reports the STEM-HAADF profile resulting from the projection of the HAADF image acquired simultaneously with the NanoCL spectrum image. Here, each QDisc can be identified as a local maximum of signal. In Fig. S2(b), the CL intensity profile is plotted in colour scale as a function of the wavelength (y-axis) and of the position along the NW axis (x-axis). This plot is obtained by adding up the spectra from the pixels having identical spatial x coordinate in the region identified by the red rectangle in fig. 2(a). In Fig. S2 (b) each QDisc appears thus as a local maximum at a given position and wavelength. The spatial coordinates of the CL peaks in Fig. S2 (b) match the QDisc positions as obtained from the HAADF profile in Fig. S2 (a). The unambiguous correlation of the emission wavelength and the emitting QDiscs can then be obtained, as demonstrated in ref []. In this previous work we pointed out that the spectra from QDiscs with minimal differences in size (one or few monolayers) yield differences in the emission energy, which can be isolated in the projected spectral image. The positions of the $1^{st}$, $2^{nd}$, $3^{rd}$ and $4^{th}$ QDiscs are indicated on the HAADF profile and in the projected spectral-spatial plot in Fig. S2(b). Their spectra are shown in Fig. 3(b), except for the $1^{st}$ one which did not emit. By fitting the surface in Fig. S2(b) with 1 peak function for each QDisc, one can get its spectral emission, spectral width, location and signal spread. The location and emission wavelength for each of the fitted QDiscs is shown in Fig. 3(b) by crosses. The average signal spread (from 5 QDiscs) is 4.9±0.9 nm.

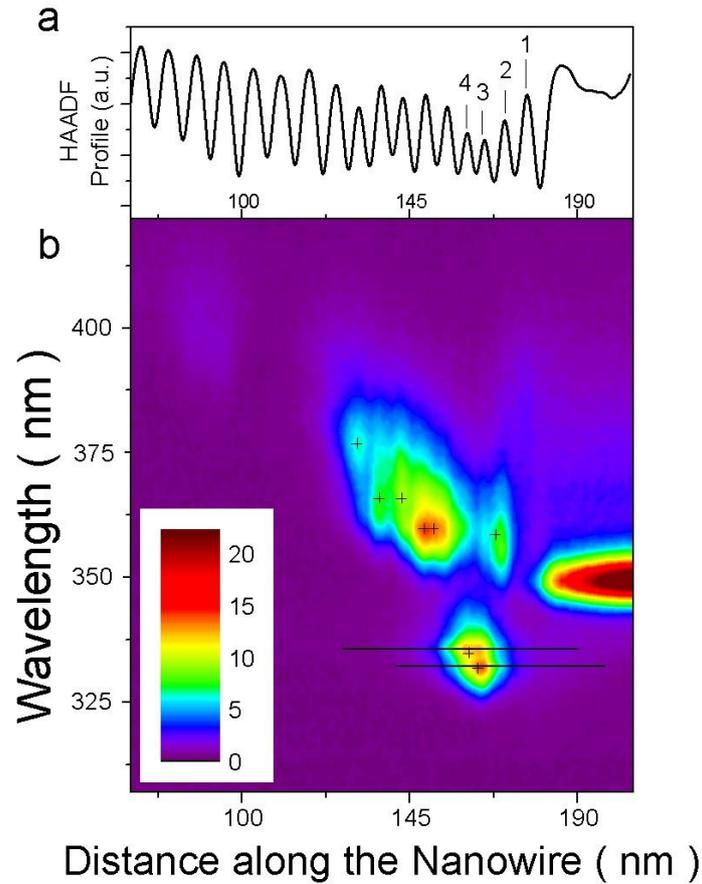

Figure S2: (a) Projected STEM-HAADF profile plotted versus the coordinate along the NW axis. The numbers indicate the positions of the first 4 QDiscs. (b) Combined spatial-spectral surface plot indicating the variation of wavelength and intensity of the emission from the NW for a range of positions along the NW axis in the region close to the GaN NW covered by AlN (shown on the right). The lines give the location of the profiles on Figure 6. The crosses indicate the centre of the QDiscs as obtained from the fitting procedure developed in ref. [19].

## References:

[1] Qian F, Li Y, Gradecak S, Wang D, Barrelet C J and Lieber C M 2004 *Nano Lett.* **4** 1975

[2] Thelander C *et al* 2006 *Materials Today* **9** 28

[3] Yan R, Gargas D, and Yang P 2009 *Nature Photonics* **3** 569

[4] Carnevale S D, Yang J, Phillips P J, Mills M J and Myers R C 2011 *Nano Letters* **11** 866

[5] Qian F, Gradecak S, Li Y and Lieber C M 2005 *Nano Letters* **5** 2287

[6] Tchernycheva M et al., 2007 *Nanotechnology* **18** 385306

[7] Rigutti L et al. 2010 *Nano Lett.* **10** 2939

[8] Pearton S J et al. *Progr. Mater. Sci.* 2010 **55** 1; Pearton S J, Kang B S and Gila B P 2008 *J. Nanosci. Nanotech*. **8** 99

[9] Songmuang R, Katsaros G, Monroy E, Spathis P, Bougerol C, Mongillo M and De Franceschi S 2010 *Nano Lett.* **10** 3545.

[10] Consonni V, Knelangen M, Jahn U, Trampert A, Geelhaar L and Riechert H, 2009 *Appl.Phys. Lett*. **95** 241910

[11] Zagonel L F *et al.* 2011 *Nano Lett.* **11** 568

[12] Bruckbauer J, Edwards P R, Wang T and Martin R W 2011 *Appl. Phys. Lett.* **98** 141908

[13] García de Abajo F J 2010 *Rev. Mod. Phys.* **82** 209

[14] Petersson A, Gustafsson A, Samuelson L, Tanaka S and Aoyagi Y 1999 *Appl. Phys. Lett.,* **74** 3513

[15] Edwards P R and Martin R W 2011 *Semicond. Sci. Technol.* **26** 064005

[16] Pennycook S J 2008 *Scanning* **30** 287

[17] Gustafsson A, Bolinsson J, Sköld N and Samuelson L 2010 *Appl. Phys. Lett*. **97** 072114

[18] Lim S K, Brewster M, Qian F, Li Y, Lieber C M and Gradecak S, 2009 *Nano Letters* **9** 3940

[19] Jahn U, Ristic J and Calleja E 2007 *Appl.Phys. Lett*. **90** 161117

[20] Jacopin G et al. 2012 *Appl Phys Express* **5** 014101

[21] Tizei L H G and Kociak M 2012 Nanotechnology **23** 175702

[22] Tourbot G et al 2012 Nanotechnology **23** 135703

[23] Songmuang R, Landré O and B. Daudin B 2007 *Appl. Phys. Lett*. **91** 251902

[24] Calarco R, Meijers R J, Debnath R K, Stoica T, Sutter E and Lu H 2007 *Nano Lett.* **7** 2248

[25] Tchernycheva M et al. 2007, Nanotechnology **18** 385306

[26] Songmuang R, Ben T, Daudin B, González D and Monroy E, 2010 *Nanotechnology* **21** 295605

[27] International patents numbers WO 2011/148073 A1 and WO 2011/148072 A1.

[28] Galopin E, Largeau L, Patriarche G, Travers L, Glas F and Harmand J C 2011 *Nanotechnology* **22** 245606

[29] Kan S, Mokari T, Rothenberg E and Banin U, 2003 *Nature materials* **2** 155

[30] Htoon H, Hollingworth J A, Malko A V, Dickerson R and Klimov V I 2003 *Appl. Phys. Lett.* **82** 4776

[31] Renard J, Kandaswamy P K, Monroy E and Gayral B, 2009 *Appl. Phys. Lett.* **95** 131903

[32] F. Furtmayr et al. 2011 *Phys. Rev. B* **84** 205303

[33] Rigutti L, Jacopin G, Largeau L, Galopin E, De Luna Bugallo A, Julien F H, Harmand J-C, Glas F and Tchernycheva M 2011 *Phys. Rev. B* **83** 155320

[34] Jacopin G, Rigutti L, Largeau L, Fortuna F, Furtmayr F, Julien F H, Eickhoff M, and Tchernycheva M, 2011 *J. Appl. Phys.* **110** 064313

[35] Reshchikov MA, Morkoç H 2005 *J. Appl. Phys.* **97** 061301

[36] Li Q, Wang GT, 2010 *Nano Lett* **10** 1554

[37] Kwon Y-H, Shee SK, Gainer GH, Park GH, Hwang SJ, and Song JJ, 2000 *Appl. Phys. Lett.*, **76** 840.

[38] Schlager JB, Bertness KA, Blanchard PT, Robins LH, Roshko A, Sanford NA, 2008 *J. Appl. Phys*. **103** 124309.

[39] S. Dhamodaran, 2011 *Optical Materials* **33** 332.

[40] Hestroffer K et al 2010 *Nanotechnology* **21** 415702

[41] J. Nelayah, et al., 2007 *Nature Physics* **3** 348

[42] Kociak M, et al., 2000 *Phys. Rev. B* **61** 13936

[43] Reed B, Chen J, MacDonald N, Silcox J, Bertsch G. 1999 *Phys. Rev. B* **60**, 5641

[44] A. Howie, S. Pennycock, 1980 Phil. Mag. A. **6** 809

[45] Baird L, Ang G H, Low C H, Haegel N M, A A, Li Q and Wang G T 2009 *Physica B* **404** 4933

[46] Yang JW et al. 1996 *Appl. Phys. Lett.* **69** 3566

[47] Duboz J Y, Binet F, Dolfi D, Laurent N, Scholz F, Off J, Sohmer, A, Briot O and Gil B 1997 *Mater. Sci. Eng.* B **50** 289

[48] Barjon J, Brault J, Daudin B, Jalabert D, Sieber B 2003 *J. Appl. Phys.* **94** 2755

[49] Baird L et al. 2011 *App. Phys. Lett.* **98** 132104